\documentclass[aps,prl,twocolumn,superscriptaddress]{revtex4-1}
\usepackage{graphicx}
\usepackage{amsmath,amsfonts,amssymb,wasysym}
\usepackage{physics}
\usepackage{color}
\newcommand{\diag}{\text{diag}}
\definecolor{myblue}{rgb}{0.368417, 0.506779, 0.709798}
\definecolor{myyellow}{rgb}{0.880722, 0.611041, 0.142051}
\definecolor{mygreen}{rgb}{0.560181, 0.691569, 0.194885}
\definecolor{myred}{rgb}{0.922526, 0.385626, 0.209179}
\begin{document}
\title{
Slowest kinetic modes revealed by metabasin renormalization
}
\author{Teruaki Okushima}
\email{okushima@isc.chubu.ac.jp}
\affiliation{
Science and Technology Section, General Education Division, College of Engineering, Chubu University, Matsumoto-cho, Kasugai, Aichi 487-8501, Japan}
\author{Tomoaki Niiyama}
\email{niyama@se.kanazawa-u.ac.jp}
\affiliation{
College of Science and Engineering,
Kanazwa University,
Kakuma-cho, Kanazawa, Ishikawa 920-1192, Japan
}
\author{Kensuke S. Ikeda}
\email{ahoo@ike-dyn.ritsumei.ac.jp}
\affiliation{
College of Science and Engineering, Ritsumeikan University, 
Noji-higashi 1-1-1, Kusatsu 525-8577, Japan}
\author{Yasushi Shimizu}
\email{shimizu@se.ritsumei.ac.jp}
\affiliation{Department of Physics, Ritsumeikan University, Noji-higashi 1-1-1, Kusatsu 525-8577, Japan}
\date{\today}
\begin{abstract} 
Understanding the slowest relaxations of complex systems, 
such as relaxation of glass-forming materials, 
diffusion in nanoclusters, 
and 
folding of biomolecules, 
is important for physics, chemistry, and biology. 
For a kinetic system, the relaxation modes are determined by diagonalizing its transition rate matrix. 
However, for realistic systems of interest, numerical diagonalization, as well as
extracting physical understanding from the diagonalization results, 
is  difficult due to the high dimensionality.
%
Here, we develop an alternative and generally applicable  method of extracting the long-time scale
relaxation dynamics by combining the metabasin analysis of Okushima \textit{et al}.\ [Phys. Rev. E {\bf 80}, 036112 (2009)] 
and a Jacobi method.
We test the method on a illustrative model of a four-funnel model,
for which we obtain a renormalized kinematic equation of much lower dimension 
sufficient for determining slow relaxation modes precisely.
%
%
%
The  method is successfully applied to the vacancy transport problem 
in ionic nanoparticles 
[Niiyama \textit{et al}.\ Chem. Phys. Lett. {\bf 654}, 52 (2016)], 
allowing a clear physical interpretation that the final relaxation 
consists of two successive, characteristic processes.
\end{abstract}
\maketitle
Recently,
dynamics of complex systems, 
such as
relaxation of glass-forming materials \cite{goldstein,stillingerWeber,stillingerWeber2,stillinger,heuer,APRV,debenedettiStillinger,sastory,DRB,DH,DH2,AFMK,glass,DSSSKG},
conformational transitions in biomolecules 
\cite{BeckerKerplus,wolynes,pande,wang,folding,kern,nucleosome}, and
rapid diffusion in nanoclusters \cite{bbskj,hhdksi,jkk,hshias,hbssh,DSSSKG,niiyama},
are being studied in a unified way 
by analyzing kinetics on rugged potential energy surfaces \cite{wales0,wales,stillingerBook}.
In the  basin hopping approach,
the  phase space is divided into  basins of minima 
on the potential energy surface, and
the local equilibrium in each basin is assumed 
to be achieved immediately. 
In this approach,
the dynamical properties are  described by 
the transition rate matrix,
which characterizes all the transitions between adjacent basins.
Hence,
the numerical diagonalization of the transition rate matrix
enables us in principle to derive  every detail of the  time evolution.
However, for realistic, complicated systems,
this  procedure is impractical because of the huge matrix dimensions.
Even if the diagonalizations were computable,
extracting physical understandings 
from the 
large number of large dimensional eigenvectors
would be very difficult.
In order to reduce the matrix dimensionality,
various coarse-graining methods,
such as lumping \cite{lumping,clumping,clumping2},
Perron-cluster  analysis \cite{BPE}, 
and discrete path sampling \cite{wales,dps}
have been developed.
Nevertheless,
it is well known that
there is as yet no coarse-graining
method applicable to
such realistic, complicated  systems without 
deterioration of the accuracy of relaxation modes and relaxation rates
\cite{BPE,wales}.

In this Rapid Communication, to overcome this difficulty, 
we develop an alternative renormalization method 
tailored for extracting the slow dynamics precisely,
which is based upon  metabasin analysis \cite{maeno,kfs} 
and a variant of the Jacobi rotation method for matrix diagonalization. 
Through the accurate renormalization procedure, 
a slow kinetic equation is generated
that can reproduce  the slow relaxation modes precisely.  
Further, 
we successfully apply the renormalization method
to elucidate the final relaxation process of 
fast vacancy transport in ionic nanoparticles, 
which was first observed experimentally by \cite{kimura}
and explored numerically by \cite{niiyama}.

In the basin hopping approach,
the kinetic state
is described by
the distribution of probability, $p_i$, of being in the basin of
$i$th local minimum (LM) for $i=1,2,\dots,n$,
where $n$ denotes the number of LMs.
The kinetic equations 
are given by
$ {d p_i}/{dt}=
\sum_{j=1}^n k_{ij}p_j-p_i\sum_{j=1}^n k_{ji}
$,
where
$k_{ij}$ 
is the transition rate from 
$j$th to $i$th LM.
In the harmonic approximation \cite{wales},
$k_{ij}$ 
is 
evaluated 
at temperature $T$,
as
$
 k_{ij}=
\nu_{ij}
\exp\left\{
{-\beta [E(\text{SP}_{ij})-E(\text{LM}_{j})]}
\right\}
$ for $i \neq j$
and 
$k_{ii}=0$,
where
$\beta=1/k_\text{B}T$
with $k_\text{B}$ Boltzmann constant.
$E(\text{LM}_{j})$ and  $E(\text{SP}_{ij})$ are
the potential energies
at
$j$th LM
and
at
the saddle point (SP) 
connecting the basins of $\text{LM}_i$ and $\text{LM}_j$, respectively.
The prefactor $\nu_{ij}$ is the frequency factor
of this transition, which is determined from the second derivatives of 
potential energy at LM$_j$ and at SP$_{ij}$.
Now,
the transition rate matrix $K$ is defined by
$(K)_{ij}=k_{ij}-\delta_{ij}\sum_{j'=1}^n k_{j' i}$
for $i,j=1,\dots,n$.
Consequently,
the kinetic equations 
can be expressed in a  matrix form:
${d\vb*{p}}/dt=K\vb*{p}$ where
$\vb*{p}=(p_1,\dots,p_n)^T$ 
with the superscript $T$ denoting the  transpose.
We assume 
the equilibrium, $\lim_{t\to\infty}\vb*{p}(t)$, to be unique.
Accordingly,
the eigenvalues of $K$ satisfy 
$
0= \lambda_0 > \lambda_1 \geqslant \dots \geqslant \lambda_{n-1}
$
\cite{haken}.
The equilibrium $\vb*{p}(\infty)$ coincides with
the zeroth eigenvector of $K$,  
and the first, second, $\dots$ eigenvectors of $K$
represent the slowest relaxation modes 
with the relaxation times of $|\lambda_1|^{-1} \geqslant |\lambda_2|^{-1}\geqslant\dots$,
respectively.

Next we consider sets of LMs, called metabasins (MBs), 
that are determined  with the use of monotonic sequences
\cite{maeno}.
A sequence $\text{LM}_{i_1}\to \text{LM}_{i_2} \to \dots$
is called monotonic if
it consists only of  most probable transitions.
Hence,
monotonic sequences with the same terminal LM 
belong to the same MB.
This classification scheme groups all $n$ LMs into a finite number, 
say $m$, of MBs: e.g.,
$\text{MB}_1 = \{\text{LM}_{\sigma(1,1)}, \dots, \text{LM}_{\sigma(1,n_1)}\},\ 
\text{MB}_2 =\{\text{LM}_{\sigma(2,1)}, \dots, \text{LM}_{\sigma(2,n_2)}\},\dots, 
\text{MB}_m = \{\text{LM}_{\sigma(m,1)}, \dots,\\   \text{LM}_{\sigma(m,n_m)}\}
$.
Here, 
$n_\ell$ denotes  the number of elements in $\text{MB}_{\ell}$
and
$\sigma(\ell,i)$  gives the index $j$ of $\text{LM}_j$
that is the $i$th energy LM in $\text{MB}_{\ell}$.
We rearrange the columns and rows of  $K$
in the ordering of
$\sigma(1,1),\dots, \sigma(1,n_1),\dots,\sigma(m,1),$
$\dots,\sigma(m,n_m) $,
and the resultant matrix is  denoted by $K_{\sigma}$.
\begin{figure}[t]
\begin{center}
\includegraphics[width=8.5cm]{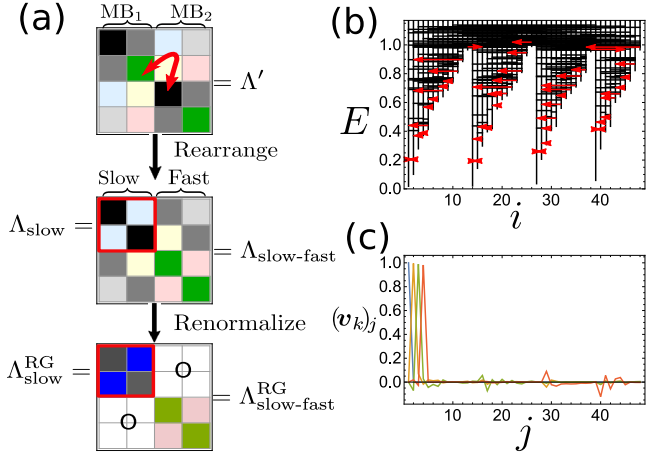} 
\end{center}
\caption{%
(a) Renormalization procedure is illustrated for a two-MB model.
Upper panel: In $\Lambda'$ representation,
the diagonal blocks are 
slow block in MB$_1$ (black),
fast block in MB$_1$ (green),
slow block in MB$_2$ (black),
fast block in MB$_2$ (green),
and off-diagonal bocks are interactions between them.
Middle panel:
$\Lambda_\text{slow-fast}$ is obtained by
exchanging the positions of slow block of MB$_2$ and fast block of MB$_1$
in $\Lambda'$.
Lower panel:
Transforming
$\Lambda_\text{slow-fast}$
by Jacobi rotation $G$
produces 
renormalized matrices $\Lambda_\text{slow-fast}^\text{RG}$,
in which the slow-fast blocks are zero, and $\Lambda_\text{slow}^\text{RG}$.
\label{fig:matplot}
(b) Saddle connectivity graph \cite{maeno}
of  four-funnel model.
The horizontal  axis represents the index $i=1,2,\dots,48$ of LM$_i$ 
and 
the vertical axis represents the potential energies of LMs and SPs.
LM$_i$ is represented by the vertical line starting at $(i,E(\text{LM}_i))$.
SP$_{ij}$ is represented by the horizontal line from  
$(i,E(\text{SP}_{ij}))$ to
$(j,E(\text{SP}_{ij}))$.
The (red) arrows represent  monotonic sequences. 
The four MBs show   funnel structures \cite{wales},
where
the typical inter-MB barrier height $\sim\!1$
and
the typical intra-MB barrier height $\sim\!0.1$.
(c)
For the $k$th slowest relaxation modes of  $k=0,1,2,3$, 
the coefficients, $(\vb*{v}_k)_j$, in the following basis
are plotted:
$j= 1,\dots,4$ represent the eigen relaxation modes of $\Lambda_\text{slow}$ 
and 
$j=5,\dots,48$ represent the fast modes of $\Lambda_\text{slow-fast}$.
We see that
$(\vb*{v}_k)_j\simeq \delta_{j,k+1}$ hold. 
The deviations from $\delta_{j,k+1}$
indicate 
both
 slow-slow mode mixing for $j=1,\dots,4$, resulting in 
the renormalization of the intra-MB slow-mode couplings, 
and
 slow-fast mode mixing for $j=5,\dots,48$.
}
\label{fig:scg}
\end{figure}

In the $K_{\sigma}$ representation,
the intra-MB, diagonal blocks
tend to be larger than the
inter-MB, off-diagonal blocks, 
i.e.,
$
\max_i \{K_{\sigma(\ell,i)\sigma(\ell,j)}\}>K_{\sigma(\ell',i')\sigma(\ell,j)}
$ 
for arbitrary $\ell'\neq\ell$ and $i'$,
since 
all LMs in a MB$_\ell$ are connected by most probable transitions.
Hence,
we regard 
the  off-diagonal blocks
as perturbations to the diagonal blocks.
Thus, we first consider 
the block diagonal matrix $\diag (K_{1},\dots,K_\ell,\dots,K_{m})$,
where 
$K_\ell$ is given by
$(K_{\ell})_{ij}=k_{\sigma(\ell,i),\sigma(\ell,j)}
-\delta_{ij}\sum^{n_\ell}_{{j'=1}}
k_{\sigma(\ell,j'),\sigma(\ell,i)}$.
Namely,
 $K_\ell$ describes the intra-MB$_\ell$ relaxations,
whose $j$th eigenvalues $\lambda_{\ell,j}$ satisfy
$0=\lambda_{\ell,0}>
\lambda_{\ell,1}\geqslant\dots 
\geqslant
\lambda_{\ell,n_\ell-1}$.
The intra-MB relaxation modes are obtained as follows:
By using the  local equilibrium $\vb*{p}_{\ell,0}$ in MB$_\ell$ satisfying 
$K_{\ell} \vb*{p}_{\ell,0}=0$,
we form
$T_{\ell}=D_\ell^{-1} K_{\ell} D_\ell$,
using the diagonal matrix $D_\ell$ with
$(D_\ell)_{i,i}=\sqrt{(\vb*{p}_{\ell,0})_i}$.
$T_{\ell}$ is 
the symmetric matrix and can be
diagonalized with
an orthogonal matrix 
$S_\ell=\left[
\sqrt{\vb*{p}_{\ell,0}}, \vb*{v}_{\ell,1},\dots,\vb*{v}_{\ell,n_\ell-1}
\right]$,
as
$S_\ell^T T_\ell S_\ell=\diag (0,\lambda_{\ell,1},\dots,\lambda_{\ell,n_\ell-1})\equiv \Lambda_\ell$,
where
the $j$th eigenvectors, $\vb*{v}_{\ell,j}$, 
describe  the $j$th intra-MB$_\ell$ relaxation modes
of relaxation rates  $\lambda_{\ell,j}$.
Note here that  $T_\ell$ 
is diagonalized  more easily than the whole system of $K_\sigma$.

Next,
we consider the inter-MB transitions.
The global equilibrium $\vb*{p}_{\text{eq}}$ satisfies $K_{\sigma}\vb*{p}_\text{eq}=0$ as well as
$\diag (K_1,K_2,\dots )\vb*{p}_\text{eq}=0$.
Hence,
the diagonal matrix $D$ with $(D)_{i,i}=\sqrt{(\vb*{p}_\text{eq})_i}$
and
$S=\diag (S_1,S_2,\dots)$
satisfy
$S^T D^{-1} \diag (K_1,K_2,\dots) D S=
\diag (\Lambda_1,\Lambda_2,\dots)$.
Hence,
the symmetric matrix $\Lambda'=S^T D^{-1}K_{\sigma}D S$
describes the couplings between  intra-MB relaxation modes.
Note that
$\Lambda'$  has nonzero off-diagonal elements
not only in inter-MB  off-diagonal blocks,
but also in intra-MB diagonal blocks [Fig.~\ref{fig:matplot}(a), upper panel].

The unperturbed fast intra-MB relaxation modes promptly decay and
would hardly contribute to the global slowest modes at all,
while
the unperturbed slow relaxation modes do interact with each other
and
mainly form 
the global slowest relaxation modes.
Hence, we introduce a certain threshold $\lambda_\text{cut}$
and divide the unperturbed relaxation modes
into two: 
the slow relaxation modes ($0\geqslant \lambda_{\ell,j}\geqslant\lambda_\text{cut}$)
and 
the fast relaxation modes ($\lambda_\text{cut}> \lambda_{\ell,j}$) 
[Fig.~\ref{fig:matplot}(a), upper panel].
For the sake of convenience,
we reorder the columns and lows of $\Lambda'$ 
in the slow-to-fast relaxation block order, 
as shown in the middle panel of Fig.~\ref{fig:matplot}(a). 
The resultant matrix is denoted by $\Lambda_\text{slow-fast}$,
where 
$\Lambda_\text{slow}$
is
the first $n_\text{slow}\times n_\text{slow}$ submatrix
with $n_\text{slow}$ denoting  the number of unperturbed slow relaxation modes.

In the following,
we first show that
the existing coarse-graining procedures for kinetic problems, 
which assume intra-MB local equilibriums,
are insufficient to obtain accurate results, as stated in  \cite{BPE}.
Then,
we develop a renormalization procedure  with the use of the Jacobi method,
where the resultant coarse-graining errors are reduced to zero.

Let us start with
exemplifying how
the coarse-graining procedure
gives rise to errors 
with use of 
the four-funnel model
depicted in 
Fig.~\ref{fig:scg}(b).
 For simplicity,
 all frequency factors, $\nu_{ij}$, in the transition rate matrix are set to be $1$.
With the use of the MB analysis,
we obtain the following four MBs:
$\text{MB}_1=\{\text{LM}_1,\dots,\text{LM}_{12}\}$,
$\text{MB}_2=\{\text{LM}_{13},\dots,\text{LM}_{26}\}$,
$\text{MB}_3=\{\text{LM}_{27},\dots,\text{LM}_{37},\text{LM}_{48}\}$, and 
$\text{MB}_4=\{\text{LM}_{38},\dots,\text{LM}_{47}\}$.
Here we set $\lambda_\text{cut}=0$ and
the slow relaxation modes are thereby composed of
 four intra-MB local equilibria ($n_\text{slow}=4$).
The corresponding
$4 \times 4$ submatrix $\Lambda_\text{slow}$ 
has
the eigenvalues of $0,\ -0.104,\ -0.208$, and $-0.355$,
which are approximations to
the  exact slowest four eigenvalues 
of  $0,\ -0.089,\ -0.154$, and  $-0.235$
at $\beta=5$.
The discrepancies come from
the inter-MB transitions.
Figure~\ref{fig:scg}(c)
shows that
the global relaxation modes 
are composed 
not only of slow unperturbed modes
but also  of fast relaxation modes.
Namely,
the couplings between slow and fast relaxation modes 
 in $\Lambda_\text{slow-fast}$
also
modify the couplings among the intra-MB slow modes.
This is the reason why
any existing coarse-graining procedures for kinetic problems, 
which simply neglect the fast intra-MB relaxation modes
and
assume the states to be linear combinations of intra-MB local equilibriums,
are insufficient to obtain accurate results.

Now we construct a renormalized transition matrix, 
$\Lambda^{\text{RG}}_{\text{slow}}$,
describing the global slowest relaxation modes accurately.
To this end,
we use a Jacobi rotation
$\Lambda_\text{slow-fast}\mapsto 
\Lambda_\text{slow-fast}^\text{RG}
=G^T \Lambda_\text{slow-fast} G$
such that
the resultant couplings  between
slow and fast modes,
 $(\Lambda_\text{slow-fast}^\text{RG})_{ij}$ 
with  $i\leqslant n_\text{slow} <j$,
are vanishing.
We here choose 
the repeated Givens matrix $G=G_1G_2\dots G_r$ for $G$,
where
$G_s=G(p_s,q_s,\theta_s)$ are defined by
$
\left(G(p,q,\theta)\right)_{pp} = 
\left(G(p,q,\theta)\right)_{qq} = \cos \theta$,
$
\left(G(p,q,\theta)\right)_{pq} = 
-\left(G(p,q,\theta)\right)_{qp} =\sin\theta $, 
$\left(G(p,q,\theta)\right)_{ii} =1 $ for $i\neq p,\ q$,
otherwise 
$\left(G(p,q,\theta)\right)_{ij} =0 $.
In actual computation, we repeat the following procedures for $s=1,2,\dots,r$:
We first choose $p_s, q_s$ randomly from $p_s\leqslant n_\text{slow}<q_s$,
and  set $\theta_s$ as
$\theta_s=\frac12 \tan^{-1}[{2 (A_{s-1})_{p_sq_s}}/
((A_{s-1})_{p_sp_s}-(A_{s-1})_{q_sq_s})]$,
so as to eliminate $(p_s,q_s)$-entry of $A_s$,
where $A_{s}=G_{s}^T\dots G_{1}^T \Lambda_\text{slow-fast} G_1\dots G_{s}$
and $A_0=\Lambda_\text{slow-fast}$.
In short,
this procedure is a Jacobi method, originally developed for symmetric matrix diagonalization \cite{matrixcomp},
which is 
modified to eliminate 
not all the off-diagonal elements, 
but only those of the  slow-fast couplings. 
Therefore, as the procedure is repeated sufficiently many times 
(say, $r$ times),
the couplings between slow and fast relaxation modes
in $A_r$ do converge to zero and
these modes 
are decoupled in the final representation.
Hence,
we set
$\Lambda_\text{slow-fast}^\text{RG}= A_r$
and
$\Lambda^\text{RG}_\text{slow}$ is defined by
the first 
$n_\text{slow}$-by-$n_\text{slow}$ submatrix of 
$\Lambda_\text{slow-fast}^\text{RG}$
[Fig.~\ref{fig:matplot}(a), lower panel].
It is $\Lambda^\text{RG}_\text{slow}$ that 
exactly describes the transitions among the renormalized slow relaxation modes.

Using  the four-funnel model, 
we examined how the renormalization procedure works.
First, we confirmed  that 
the slow-fast coupling elements of 
$\Lambda^\text{RG}_\text{slow-fast}$ 
do converge to zero
as in the lower panel of Fig.~\ref{fig:matplot}(a). 
The resultant matrices 
$\Lambda_\text{slow}$ 
and 
$\Lambda^\text{RG}_\text{slow}$
are as follows:
%
\begin{eqnarray*}
\Lambda_\text{slow}&=&
\left(
\begin{smallmatrix}
-0.108& 0.078& 0.019& 0.020\\ 
0.078& -0.142&   0.037& 0.033\\
0.019& 0.037& -0.185&   0.144\\
0.020& 0.033& 0.144& -0.232
\end{smallmatrix}
\right),\\
\Lambda^\text{RG}_\text{slow}&=&
\left(
\begin{smallmatrix}
-0.088& 0.057& 0.018& 0.021\\
0.057& -0.104&   0.028& 0.023\\
0.018& 0.028& -0.128&  0.090\\
0.021& 0.023& 0.090& -0.159  
\end{smallmatrix}
\right).
\end{eqnarray*}
Comparing these matrices,
we see that
the coupling terms between slow modes 
are modified
by relative ratios 
of $0.01$--$0.1$,
as  a result of the renormalization.
Due to the renormalization effect,
we get the right eigenvalues of
$
0,\
-0.089,\ -0.154$, and $-0.235$
by diagonalizing $A^\text{RG}_\text{slow}$,
which 
numerically 
agree
with the above mentioned exact values of the slowest four eigenvalues at  $\beta=5$. 

Finally,
the kinetics of vacancy diffusion in KCl nanoclusters \cite{niiyama}
is examined for a realistic problem.
Suppose
one chlorine ion is extracted from a cube of ionic crystal,
with equal $N_L$-atom edges.
Assume also that
$N_L$ is an odd number $2 n_L+1$
and 
the resultant (${N_L}^3-1$)-atom cluster
is electrically neutral.
Then,
the vacancy moves around the cluster,
which induces atomic diffusion.
Note that
the cubic form  of the cluster is kept in the course of time evolution,
when the temperature is sufficiently low \cite{cubic}.
At such low temperatures,
the position of the vacancy is specified
by the cubic lattice point $(n_x,n_y,n_z)$ with 
$-n_L \leqslant n_x,n_y,n_z \leqslant n_L$.
In addition,
we are able to find the atomic structure of LM specified by $(n_x,n_y,n_z)$
as follows:
First,
atoms are arranged at
$d(m_x,m_y,m_z)$ with lattice constant $d=3.147\ \text{\AA}$ for KCl, 
where $(m_x,m_y,m_z)\neq (n_x,n_y,n_z)$ and $-n_L \leqslant m_x,m_y,m_z \leqslant n_L$.
Then, the  configuration of atoms
is relaxed to the LM energy structure by, 
e.g.,  steepest descent method.
In this way,
the LM atomic structure is 
assigned to  $(n_x,n_y,n_z)$.
For computational details of enumerating LMs as well as  SPs,
we refer the reader to Ref.~\cite{niiyama}. 
\begin{figure}[t]
\includegraphics[width=8.cm]{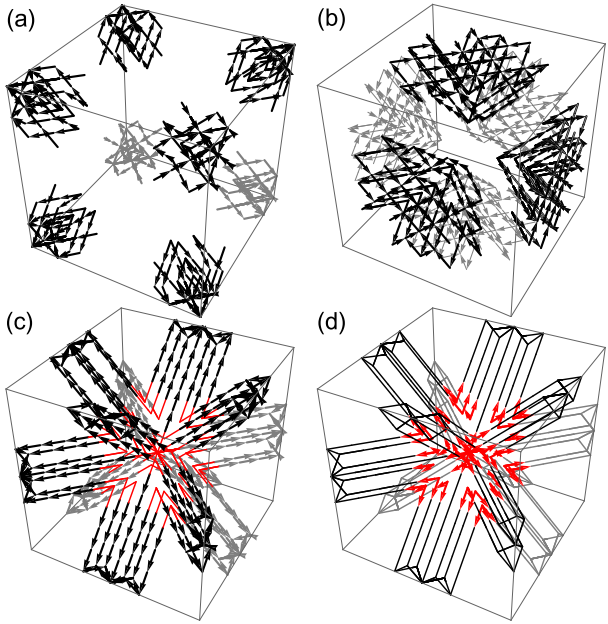}
\caption{
MBs of the $N_L=13$ cluster 
are represented by arrows  (See text):
(a) Eight MBs located at the vertexes,
(b) Six MBs located at the faces,
(c) 12 MBs located at the edges,
and
(d) 9 saddlelike MBs located in the central part.
Thin red lines in (c) and thin black lines in (d) are drawn 
to show that the saddlelike MBs are hubs among the edge MBs.
}
\label{fig:mb}
\end{figure}

The MBs of the $N_L=13$ cluster
at temperature $k_\text{B} T=0.03$ eV
are depicted in  Fig.~\ref{fig:mb},
where
the monotonic sequences,
$\text{LM}_{i_1}\to\text{LM}_{i_2}\to\dots$, 
are shown by
the arrows,
 $(n_x,n_y,n_z)_{i_1} \to
(n_x,n_y,n_z)_{i_2} \to\dots$, 
which connect the corresponding vacancy lattice points.
The collections of LMs with the same terminal LMs 
represent  MBs.
In Fig.~\ref{fig:mb}(a),
the eight most stable MBs, 
containing
the lowest energy terminal LMs of $(\pm n_L,\pm n_L,\pm n_L)$,
are shown.
In addition,
there exist  six MBs
with terminal LMs at  the centers of faces 
$(\pm n_L,0,0)$, $(0,\pm n_L,0)$, and $(0,0,\pm n_L)$ [Fig.~\ref{fig:mb}(b)],
and 
 12 MBs
with 
terminal LMs at the
centers of edges 
$(\pm n_L,\pm n_L,0)$,
$(\pm n_L,0,\pm n_L)$, and
$(0,\pm n_L,\pm n_L)$ [Fig.~\ref{fig:mb}(c)].
Moreover,
due to the cubic symmetry, 
there are ``saddlelike'' LMs, 
which have at least  two monotonic sequences reaching 
different terminal LMs.
For example,
eight monotonic sequences emanating from $(0,0,0)$ 
have terminal LMs at  $(\pm n_L,\pm n_L,\pm n_L)$. 
Hence,
$(0,0,0)$ mediates the transitions among the vertex MBs like a saddle.
To obtain a more coarse-grained description,
we apply the MB analysis again 
only to saddlelike LMs and the SPs connecting these LMs \cite{endnote}, 
to classify them into nine saddlelike MBs,
as shown in Fig.~\ref{fig:mb}(d).

\begin{figure}[t]
\includegraphics[width=8cm]{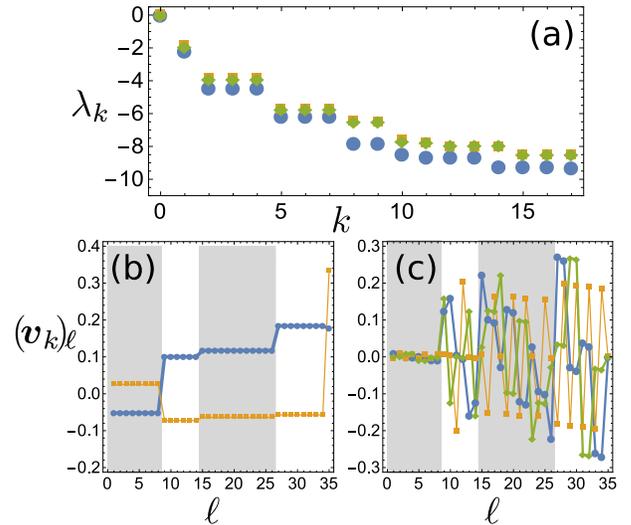}
\caption{
(a) For the $N_L=13$ cluster at $k_\text{B} T=0.03$ eV,
the slowest relaxation rates, 
$\lambda_k$ [$\times 10^{5}$s$^{-1}$], 
are plotted
as a function of  $k=0,1,\dots ,17$.
Markers 
{\color{myblue}{\CIRCLE}},
{\color{mygreen}{$\blacklozenge$}}, and
{\color{myyellow}{$\blacksquare$}}
indicate
the slowest eigenvalues of
$\Lambda_\text{slow}$,
$\Lambda_\text{slow-fast}$, and 
$\Lambda^\text{RG}_\text{slow}$,
respectively.
For the $k$th relaxation modes,
$(\vb*{v}_k)_\ell$
are plotted
in (b) and (c).
In the horizontal axes, 
$\ell=1,\dots,8$ (shaded) correspond to the vertex MBs,
$\ell=9,\dots,14$ to the face-centered MBs, 
$\ell=15,\dots,26$ (shaded) to the edge-centered MBs, and
$\ell=27,\dots,35$ to the saddlelike MBs. 
(b) Inter hetero-MB relaxation modes:
{\color{myblue}{\CIRCLE}} and 
{\color{myyellow}{$\blacksquare$}}
are the results of $k=1$ and $11$, respectively,
where
$(\vb*{v}_k)_\ell =(\vb*{v}_k)_{\ell'} $ 
hold
for $\ell, \ell'$   in the same type of  MBs.
In this case,
the equilibrations occur only among the {\em different} types of MBs.
(c) Inter iso-MB relaxation modes:
${\color{myblue}{\CIRCLE}}$,
${\color{myyellow}{\blacksquare}}$,
and 
${\color{mygreen}{\blacklozenge}}$
are, respectively,
the results of 
$k=2$, 3, and 4,
where
$\sum_{\ell \in \text{same type of MBs}} (\vb*{v}_k)_\ell \simeq 0$ hold.
In this case,
the equilibrations can occur only among the same type of MBs.
}
\label{fig:kekka}
\label{fig:kekka2}
\end{figure}

We divide the intra-MB relaxation modes
into slow and fast modes by setting 
$\lambda_\text{cut}=5.0\times 10^{5}\ \text{s}^{-1}$.
The resultant total dimension of slow modes is 
$n_\text{slow}=137$.
We first diagonalize
the $137\times 137$ dimensional $\Lambda_\text{slow}$.
The eigenvalues are plotted in
Fig.~\ref{fig:kekka}(a),
where
the approximate result  is
in qualitative agreement with
the exact result of $\Lambda_\text{slow-fast}$,
although
 $n_\text{slow}=137$
is a quite small dimension compared to the full dimension of 1099.
We then apply
the renormalization procedure developed above
to $\Lambda_\text{slow-fast}$,
and obtain
the renormalized
$\Lambda^\text{RG}_{\text{slow}}$
and 
the Givens matrix $G$.
After diagonalizing 
$\Lambda^\text{RG}_{\text{slow}}$,
we also plot the eigenvalues of 
$\Lambda^\text{RG}_{\text{slow}}$
in Fig.~\ref{fig:kekka}(a),
which shows that
the slowest relaxations
are exactly described
by
the quite small $137\times137$ matrix
of 
$\Lambda^{\text{RG}}_{\text{slow}}$.

Now lastly,
we show the usefulness of 
the metabasin representation
for describing the slowest relaxation modes. 
We plot
the intra-MB$_\ell$ equilibrium components, 
$(\vb*{v}_k)_\ell$
of the $k$th slowest relaxation modes, $\vb*{v}_k=G\vb*{v}_k^\text{RG}$, 
in Figs.~\ref{fig:kekka2}(b) and \ref{fig:kekka2}(c),
from which we see that the global relaxations  are grouped into two types:
inter hetero-MB relaxation modes 
and 
inter iso-MB relaxation modes. 
As shown in Fig.~\ref{fig:kekka2}(b),
the inter hetero-MB  modes
equilibrate the disturbance only among  different types of MBs.
As a result,
they equilibrate the disturbance along the radial direction from the cubic center.
The plot for $\vb*{v}_1$ in Fig.~\ref{fig:kekka2}(b)
shows that
the bottleneck of equilibration 
is the process transporting the vacancy to the vertex MBs.
On the other hand,
the iso-MB equilibration modes equilibrate just among the same type of MBs,
and moreover 
typically localize,
as shown in Fig.~\ref{fig:kekka2}(c).
For example,
as depicted in Fig.~\ref{fig:kekka2}(c),
the vertex MBs hardly equilibrate at all
in these  modes. 
To sum up,
the slowest kinetics 
is  
 two-step relaxation,
the inter iso-MB relaxations of $\lambda_2$, $\lambda_3$, and $\lambda_4$,
followed
by the slowest inter hetero-MB relaxation of  $\lambda_1$.
It should be noted that
these results were obtained  
with the use of 
the high accuracy renormalization procedure
combined with the analysis by MB representation.
Our method provides a firm and systematic basis for the elucidation in \cite{niiyama},
where the bottleneck in the mixing process of the KCl cluster was numerically studied 
with the use of mean first passage times \cite{mfpt} 
from the center LM to the vertex LMs.

In summary, 
we developed a  renormalization procedure 
for transition rate matrices based on metabasin analysis, 
which is an accurate and efficient method 
for computing slowest relaxation modes. 
%
We also show,
with the use of the 
multifunnel model
and the ionic nanoparticle diffusion model, 
 that 
the metabasin analysis is  useful 
for grasping when, where, and how  global equilibration occurs.
Finally,
it should be noted that this procedure can be extended to be applicable 
to transition probability matrices of 
discrete-time kinetic equations with small modifications \cite{sm}. 
We hope that
with these methods,
characteristics  of slowest relaxations 
are revealed for generic multi-metabasin systems.

\begin{acknowledgments}
Y. S. and T. O.
are supported by 
Grant-in-Aid for challenging Exploratory Research
(Grant No. JP15K13539) 
from the Japan Society for the Promotion of Science. 
T. O. 
expresses gratitude to Kiyofumi Okushima and Naoto Sakae
for 
enlightening discussions and continuous encouragement.
The authors are very grateful to Shoji Tsuji and Kankikai 
for the use of their facilities at Kawaraya during this study.

\end{acknowledgments}


\begin{thebibliography}{99}
\bibitem{endnote}
Note here that saddlelike LMs are not saddles but minima. 
Hence, the coarse-graining procedures 
for saddlelike LMs are the same as those of usual LMs.
\end{thebibliography}
\end{document}